\documentclass{article}

     \usepackage[final]{neurips_2019}

\usepackage[utf8]{inputenc} 
\usepackage[T1]{fontenc}    
\usepackage{hyperref}       
\usepackage{url}            
\usepackage{booktabs}       
\usepackage{amsfonts}       
\usepackage{nicefrac}       
\usepackage{microtype}      
\usepackage{graphicx}
\usepackage{enumitem}
\usepackage{array,multirow}
\usepackage{media9}
\usepackage{hyperref}
\usepackage{amsmath}

\usepackage{multirow}
\usepackage{tabularx}
\usepackage{subfig}
\newcolumntype{Y}{>{\centering\arraybackslash}X}
\usepackage[export]{adjustbox}
\newcommand\norm[1]{\left\lVert#1\right\rVert}

\hypersetup{
    colorlinks=true,
    linkcolor=black,
    filecolor=black,      
    citecolor=black,
    urlcolor=cyan
}

\title{A Study into Echocardiography View Conversion}

\author{%
  Amir H. Abdi$^1$~\thanks{Joint first authors; the order is selected randomly.}
  \And
  Mohammad H. Jafari$^1$~\footnotemark[1] 
  \AND
  Sidney Fels$^1$ \And
  Theresa Tsang$^{12}$~\thanks{Joint senior authors} \And
    Purang Abolmaesumi$^1$~\footnotemark[2]  \\
  \\
  $^1$ Electrical and Computer Engineering Department, University of British Columbia, Vancouver, Canada
  \\$^2$  Vancouver General Hospital, Vancouver, Canada
}

\begin{document}

\maketitle

\begin{abstract}
Transthoracic echo is one of the most common means of cardiac studies in the clinical routines. During the echo exam, the sonographer captures a set of standard cross sections (echo views) of the heart. Each 2D echo view cuts through the  3D cardiac geometry 
via a unique plane. 
Consequently, different views share some limited information.
In this work, we investigate the feasibility of generating a 2D echo view using another view based on adversarial generative models.
The objective optimized to train the view-conversion model is based on the ideas introduced by LSGAN, PatchGAN and Conditional GAN (cGAN).
The size and length of the left ventricle in the generated target echo view is compared against that of the target ground-truth to assess the validity of the echo view conversion.
Results show that there is a correlation of 0.50 between the LV areas and 0.49 between the LV lengths of the generated target frames and the real target frames.
\end{abstract}

\section{Introduction}

Echocardiography (echo), or the ultrasound of the heart, is the most utilized means of evaluating cardiac conditions.
There are 12 standard echo views and a multitude of them are obtained during a conventional 2D echo exam. 
Each view captures a different cardiac cross section. 
While every echo view is special and contains unique information, they are not completely independent as they are projections of the same 3D structure.

The  ill-posed problem of 3D reconstruction from abstract representations has recently gained some traction with the advent of convolutional neural networks~\cite{klokov2019probabilistic,xie2019pix2vox}.
Inspired by the recent works in 3D reconstruction based on multiple (or single) 2D images or silhouette~\cite{han2019image}, and image domain transfer~\cite{zhu2017toward,hoffman2017cycada},
we hypothesize that the complete 3D cardiac structure can be reconstructed given a certain number of echo views.
Assuming this unproven hypothesis to be correct,
various cross sections of the reconstructed 3D heart can be potentially estimated to create the rest of the echo views.

With the above intuition, 
and considering the fact that
2D echo views share some features being cross sections of the same cardiac structures,
we are eager to understand whether a mapping can be learned from an echo view to another. In other words, we investigate the feasibility of generating the heart structure at a target view, given another cross section with the view of the heart from a different angle.
This work is a preliminary study which investigates the feasibility of transforming between two of the most-utilized standard echo views, namely the apical two-chamber (2CH) and the apical four-chamber (4CH). 
The 4CH view captures the horizontal long axis of the heart (\emph{i.e.} a four-chamber view), while the 2CH view captures the vertical long axis (left ventricle and atrium).

\section{Method}

\subsection{Data}

The CAMUS dataset~\cite{CAMUS}, organized and released by the University of Lyon, France, is a public dataset consisting of 2CH and 4CH echo views.
In CAMUS, the end-systolic (ES) and end-diastolic (ED) frames of all echo cine loops are identified. Moreover, the  cardiac chambers of these frames are manually segmented. 
For sake of reproducibility, we limit our experiments to this publicly released dataset.
Here, we only focus on converting the ED frame of 2CH to the ED frame of 4CH.
Among the total of 450 subjects in the CAMUS dataset, 75\% (342 cases) were used for training, 5\% (18 cases) for validation, and 20\% (90 cases) for testing.

\subsection{Orthogonal Views}

The cross sections captured by the 4CH and the 2CH echo views are almost orthogonal;
therefore, 
they share information only along a single shared axis, \emph{i.e.}, the left ventricle (LV) long axis.
Given that both of these views slice through the long axis of LV, the height of LV in 2CH and 4CH of the same patient is expected to be the same.
The expert LV segmentation of the ED frame for the 2CH and 4CH views are available in the CAMUS dataset; therefore, during training, 
the source images were resized to match their LV heights to that of their corresponding target view image.
This will ensure that different zooming and machine configurations, as well as operator-induced errors, do not adversely affect the view conversion.

\subsection{GAN-enhanced Deterministic Mapping}

A conditional GAN (cGAN) is used to transform the echo frame from the source echo view to the target echo view.
The CAMUS dataset contains a single 2CH and 4CH echo cine loops for each patient. Consequently, in this study, the ill-posed problem of orthogonal echo view transformation cannot be probabilistically modeled. 
Due to this limitation, the mapping is learned with no induced noise and in a deterministic fashion.

With the $l1$ norm as the reconstruction loss, the final objective function used to train the echo view transformation model is a combination of the cGAN and LSGAN~\cite{LSGAN}, summarized as:
\begin{equation}
    D^* =  \texttt{arg}~ \underset{D}{\mathrm{min}}~ 
    ~ 
    \frac{1}{2}
    \mathbb{E}_{x\sim p(x)} 
    \big[ (D(x)-1)^2 \big] 
    + 
    \frac{1}{2}
    \mathbb{E}_{c\sim p(c)} 
    \big[ D(G(c))^2 \big] ~,
\label{eq:d}
\end{equation}
\begin{equation}
    G^* =  \texttt{arg}~ \underset{G}{\mathrm{min}}~ 
    ~ \mathbb{E}_{c\sim p(c)} 
    \big[ (D(G(c))-1)^2 \big] 
    + \mathbb{E}_{c,x\sim p(c,x)} 
    \norm{x - G(c)}_1 ~.
\label{eq:g}
\end{equation}
Here, $c$ is the condition image (source view) and $x$ is the generated image (target view).
In Eq.~(\ref{eq:g}), $\norm{x - G(c)}_1$ denotes the mean average error between the generated image and corresponding target view available in the dataset.

Given that the speckle noise distribution in echocardiography images is highly dependent on the underlying local tissue structure, 
the appearance of the noise could not be enforced with
a discriminator that considered the entire image.
We arrived at a workaround to use a patch-based discriminator~\cite{pix2pix} which separately evaluated the realness of independent small patches in the image.
Here, patch sizes of $16\times16$ were used for  images of size $256\times256$.

\section{Evaluation and Results}
The implementation, trained segmentation model, and configuration files to reproduce the current experiments are publicly shared here: \url{https://github.com/amir-abdi/echo-view2view}.

\subsection{Segmentation Model}

To evaluate the applicability of the view conversion in a  clinical setting, 
and given the fact that the two views in question (2CH and 4CH) share information only along a single dimension,
we evaluated the shape similarity of the LV between the generated target view (4CH) and the ground-truth target view of the same patient.  To this end, we automatically segment the LV region in the generated 4CH view and the real 4CH view. 

For the segmentation, 
we adapt a U-Net~\cite{ronneberger2015u} architecture trained to segment the LV region in the target view, \emph{i.e.} the 4CH,  in the ED frame. The U-net has seven down-sampling and reverse up-sampling layers, with 32 base filters doubling after each down-sampling step, to the maximum of 256 filters. The skip connections are used to concatenate the corresponding feature maps from the down-sampling route to the expansive path. 
All middle layers have convolutional filters of size $3 \times  3$, followed by batch normalization and ReLU activations. The activation in the last layer is selected to be sigmoid.

\begin{figure}[t]
 \centering
 \tiny
  \begin{tabularx}{0.99\textwidth}{@{}YYYYY@{}}
 \subfloat{\includegraphics[width=0.19\textwidth,valign=c]{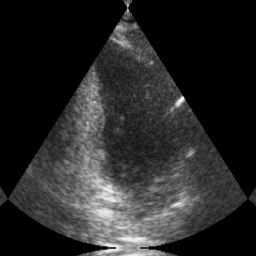}} & \subfloat{\includegraphics[width=0.19\textwidth,valign=c]{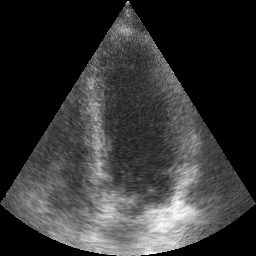}}&
\subfloat{\includegraphics[width=0.19\textwidth,valign=c]{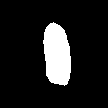}}&
\subfloat{\includegraphics[width=0.19\textwidth,valign=c]{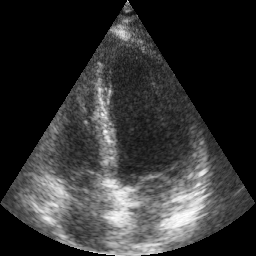}} &
\subfloat{\includegraphics[width=0.19\textwidth,valign=c]{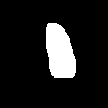}}\\[-0.3cm]

\subfloat{\includegraphics[width=0.19\textwidth,valign=c]{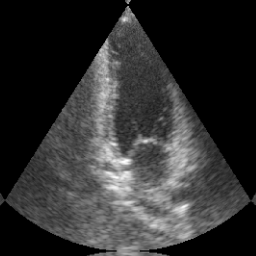}} &
\subfloat{\includegraphics[width=0.19\textwidth,valign=c]{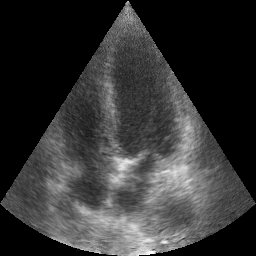}}&
\subfloat{\includegraphics[width=0.19\textwidth,valign=c]{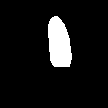}}&
\subfloat{\includegraphics[width=0.19\textwidth,valign=c]{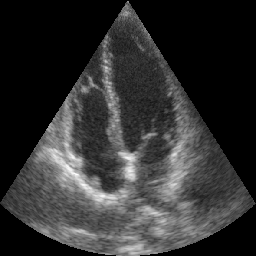}} &
\subfloat{\includegraphics[width=0.19\textwidth,valign=c]{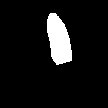}}
\\[-0.3cm]

\subfloat{\includegraphics[width=0.19\textwidth,valign=c]{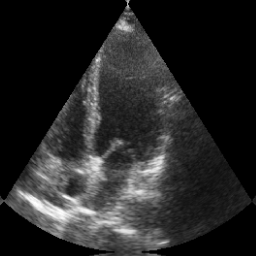}} &
\subfloat{\includegraphics[width=0.19\textwidth,valign=c]{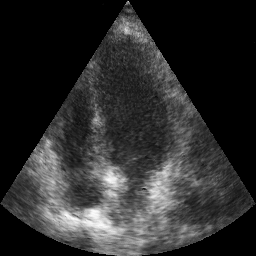}}&
\subfloat{\includegraphics[width=0.19\textwidth,valign=c]{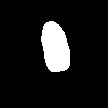}}&
\subfloat{\includegraphics[width=0.19\textwidth,valign=c]{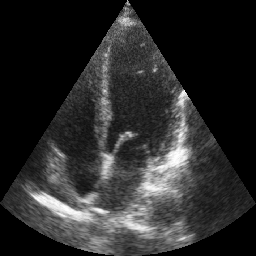}} &
\subfloat{\includegraphics[width=0.19\textwidth,valign=c]{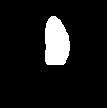}}
\\ \\
a) Source 2CH  & b) Generated 4CH  & c)  LV segmentation of the generated 4CH & d) Real 4CH & e) LV segmentation of the real 4CH 
\end{tabularx}
  \caption{Visual results for echo view conversion for three patients of the test set.
  The apical 4CH view (b) of the heart is generated from the 2CH view (a) of the same patient. 
  Notice how the height of the LV segmentation in the generated 4CH (c) and the real target 4CH (e) match. 
  }
  \label{fig1:conversion_results}
\end{figure}


\subsection{Results}

The segmented LV chambers of the generated 4CH views were compared with their counterparts in the ground-truth 4CH views in terms of  their areas and lengths along the cardiac long axis.  Let $A$ and $L$ denote the area and the length, respectively. Suppose $\hat{x}$ is the generated view and $x$ is the ground-truth view.
The results show that there is a noticeable positive correlation between the LV areas ($\text{corr}(L(\hat{x}_{LV}), L(x_{LV}))=0.50$) and lengths ($\text{corr}(L(\hat{x}_{LV}), L(x_{LV})) = 0.49$) 
of the generated and real 4CH images. 
Also, the relative error in the length of the left ventricles (12\%) is less than half of the relative area error (26\%). 
This is due to the fact that the 4CH and 2CH views do not share information along the lateral dimension of the four-chamber plane.

Although the model  learned a meaningful correlation between the two cross sections, 
it inevitably transferred irrelevant features from the source (condition) image to the generated image such as
the image contrast, speckle level, and image quality.
This is due to the fact that operators do not change machine parameters in between subsequent view collections of the same patient. Therefore, there is a positive correlation between the mentioned features of the 4CH and 2CH images in the CAMUS dataset.
This issue can be mitigated by 
conditioning the view generation on a more abstract representation of the source view.

\paragraph{Acknowledgement}
This work was conducted thanks to the funding from the Canadian Institutes of Health Research (CIHR). 

\bibliography{mybib}
\bibliographystyle{plain}

\end{document}